\documentclass[12pt]{article}
\catcode`\@=11
\@addtoreset{equation}{section}

\global\arraycolsep=2pt
\usepackage{mathrsfs,amsbsy,amssymb,latexsym,amsfonts,amsmath,cite}
\usepackage{graphicx,color}
\usepackage{physics}
\usepackage{mathtools}
\usepackage{ulem}
\usepackage{caption}
\usepackage{here,comment}

\usepackage{mathrsfs,amsbsy,amssymb,latexsym,amsfonts,amsmath,cite,bm}
\usepackage{graphicx,color}
\usepackage{mathtools}
\usepackage{enumerate}

\usepackage{xcolor}
\usepackage[pdftex,
bookmarks=true,%
bookmarksnumbered=true,%
colorlinks=flase,%
linkbordercolor={0.4 0.9 0.7},
citebordercolor={1 0.5 0},%
linkcolor=black,%
citecolor=blue,%
filecolor=black,%
pagecolor=red,%
urlcolor=blue%
]{hyperref}

\DeclareFontFamily{U}{BOONDOX-calo}{\skewchar\font=45 }
\DeclareFontShape{U}{BOONDOX-calo}{m}{n}{
  <-> s*[1.05] BOONDOX-r-calo}{}
\DeclareFontShape{U}{BOONDOX-calo}{b}{n}{
  <-> s*[1.05] BOONDOX-b-calo}{}
\DeclareMathAlphabet{\mathcalboondox}{U}{BOONDOX-calo}{m}{n}
\SetMathAlphabet{\mathcalboondox}{bold}{U}{BOONDOX-calo}{b}{n}
\DeclareMathAlphabet{\mathbcalboondox}{U}{BOONDOX-calo}{b}{n}

\newcommand{\no}{\nonumber}

\newcommand{\cH}{\mathcal H}

\newcommand{\cM}{\mathcal M}
\newcommand{\cO}{\mathcal O}

\newcommand{\cZ}{\mathcal Z}

\allowdisplaybreaks

\begin{document}

\renewcommand{\thefootnote}{\fnsymbol{footnote}}

\begin{flushright}
KUNS-2983
\end{flushright}
\vspace*{0.5cm}

\begin{center}
{\Large \bf  Remarks on effects of projective phase on\\ eigenstate thermalization hypothesis 
}
\vspace*{2cm} \\
{\large  Osamu Fukushima$^{\sharp}$\footnote{E-mail:~osamu.f@gauge.scphys.kyoto-u.ac.jp}}
\end{center}

\vspace*{0.4cm}

\begin{center}
$^{\sharp}${\it Department of Physics, Kyoto University, Kyoto 606-8502, Japan.}
\end{center}

\vspace{2cm}

\begin{abstract}
The existence of $p$-form symmetry in $(d+1)$-dimensional quantum field is known to always lead to the breakdown of the eigenstate thermalization hypothesis (ETH) for certain $(d-p)$-dimensional operators other than symmetry operators under some assumptions.
The assumptions include the mixing of symmetry sectors within a given energy shell, which is rather challenging to verify because it requires information on the eigenstates in the middle of the spectrum.
We reconsider this assumption from the viewpoint of projective representations to avoid this difficulty.
In the case of $\mathbb{Z}_N$ symmetries, we can circumvent the difficulty by considering $\mathbb{Z}_N\times\mathbb{Z}_N$-symmetric theories with nontrivial projective phases, and perturbing the Hamiltonian while preserving one of the $\mathbb{Z}_N$ symmetries of our interest.
We also perform numerical analyses for $(1+1)$-dimensional spin chains and the $(2+1)$-dimensional $\mathbb{Z}_2$ lattice gauge theory.
\end{abstract}

\setcounter{footnote}{0}
\setcounter{page}{0}
\thispagestyle{empty}

\newpage

\tableofcontents

\renewcommand\thefootnote{\arabic{footnote}}

\section{Introduction}

In isolated quantum systems, the eigenstate thermalization hypothesis (ETH) \cite{deutsch1991quantum, Srednicki:1994mfb, rigol2008thermalization} is a successful framework to explore stationary states after long-time quantum dynamics.
It provides a sufficient condition for thermalization, and has been studied in various contexts including condensed matter \cite{rigol2008thermalization,DAlessio:2015qtq, mori2018thermalization} and high energy physics \cite{Liska:2022vrd,deBoer:2016bov,Basu:2017kzo,Datta:2019jeo,Fitzpatrick:2015zha,Besken:2019bsu,Dymarsky:2019etq,Lashkari:2016vgj}.
The statement of the ETH can be phrased as follows: individual energy eigenstates of the system are thermal in a sense that the expectation values of observables are equal to their thermal ensemble average.
It is known that the ETH holds for various nonintegrable systems without symmetries \cite{rigol2008thermalization,santos2010onset,ikeda2011eigenstate,steinigeweg2013eigenstate, kim2014testing, beugeling2014finite, steinigeweg2014pushing, alba2015eigenstate, beugeling2015off, mondaini2017eigenstate, nation2018off, hamazaki2019random, khaymovich2019eigenstate,yoshizawa2018numerical, jansen2019eigenstate, Sugimoto:2020nnw,neumann1929beweis, goldstein2010long, goldstein2010approach, goldstein2010normal, reimann2015generalization,hamazaki2018atypicality,sugimoto2023rigorous}, while the existence of local conserved quantities, which is typically associated with continuous symmetries or integrability, is expected to result in its violation for many observables \cite{cassidy2011generalized, Hamazaki:2015ied, Noh:2022ijv}.

\medskip

Recently, the effects of $p$-form symmetry on the ETH is pointed out in \cite{Fukushima:2023swb}.
In general, $p$-form symmetries are characterized by $(d-p)$-dimensional topological symmetry operators in $(d+1)$-dimensional quantum field theories \cite{Kapustin:2013uxa, Kapustin:2014gua, Gaiotto:2014kfa} (for recent review, see \cite{McGreevy:2022oyu, Gomes:2023ahz, Schafer-Nameki:2023jdn, Brennan:2023mmt, Bhardwaj:2023kri, Luo:2023ive, Shao:2023gho}).
Under some reasonable assumptions, the system with $p$-form symmetry is shown to accommodate many $(d-p)$-dimensional ETH-violating observables other than the symmetry operator itself.
The assumptions consist of i) the endability of the symmetry operator, ii) mixture of symmetry sectors in a given energy shell, and iii) nonvanishing microcanonical average of the operator of our interest.
The outcome of these conditions is applicable to general nondegenerate Hamiltonians with $p$-form symmetry.
Surprisingly, this statement holds even for the systems only with nonlocal conserved quantity originated from discrete symmetries.
In many cases, it is believed that systems with local conserved quantities such as integrable models satisfy the ETH only if all of the symmetry sectors for the conserved quantities are resolved, while the ETH does not hold with the mixed symmetry sector.
The result in \cite{Fukushima:2023swb} indeed provides a proof to the violation of the ETH in the presence of higher-form symmetry.
Note that it is also applicable even to the case of non-local conserved quantities since discrete symmetries typically give them.

\medskip

Higher-form symmetries occasionally appear with projective representations rather than standard linear representations.
A projective representation is defined as a group homomorphism from a group $G$ to $\operatorname{End}(\cH)/(\mathbb{C}\backslash\{0\})$, where $\cH$ is the Hilbert space of the system.
Such a structure can be realized, for example, through higher-group, symmetry fractionalization, and 't Hooft anomaly.
The 't Hooft anomaly is defined as an obstruction to promote global symmetry to local gauge symmetry and is known to constrain the infrared theories of systems with conventional symmetries \cite{tHooft:1979rat, Kapustin:2014lwa, Kapustin:2014zva, Niemi:1983rq, Redlich:1983kn, Redlich:1983dv} or generalized symmetries \cite{Kapustin:2013uxa, Kapustin:2014gua, Gaiotto:2014kfa, Gaiotto:2017yup, Cordova:2019bsd}.
One of the significant consequences of 't Hooft anomaly for discrete symmetries is degeneracies of the ground states. 
From the viewpoint of projective representation, we can see that the degeneracy exists not only for the ground states but for all the energy eigenstates \cite{Kikuchi:2017pcp}.

\medskip

The purpose of this paper is to reconsider the sufficient conditions for the ETH-breakdown by $p$-form symmetry.
Specifically, the condition ii) above involves detailed information about the eigenstates in the middle of the energy spectrum, and thus it is rather challenging to verify this condition without explicit numerical calculations in general.
The main idea here is to work with the projective representations involving the $\mathbb{Z}_N$ $p$-form symmetry under consideration.
In this paper, we employ a $G=\mathbb{Z}_N\times \mathbb{Z}_N$ symmetry with a projective representation, and perturb the Hamiltonian to explicitly break one of the $\mathbb{Z}_N$ symmetries.
By choosing the perturbation parameter $\lambda$ appropriately, we obtain a theory with the $\mathbb{Z}_N$ $p$-form symmetry, which satisfies the condition ii).
Since the broken symmetry is expected not to affect the thermal ensemble in the thermodynamic limit, the system just reduces to have the $(d-p)$-dimensional ETH-violating operators eventually.
It is remarkable that along this construction, we do not need any direct reference to the details of the energy eigenstates in the middle of the spectrum.

\medskip

This paper is organized as follows. 
Section \ref{sec:p-ETH} provides the brief review of the ETH-violation caused by $p$-form symmetry.
In Section \ref{sec:Projective}, we discuss the degeneracies caused by a $\mathbb{Z}_N\times\mathbb{Z}_N$ projective representation, and present a symmetry-breaking perturbation so that the conditions for the ETH-violation are satisfied.
We also carry out numerical analysis to demonstrate this argument in Section \ref{sec:demonstration}.
The models includes $(1+1)$-dimensional $\mathbb{Z}_2$-symmetric/$\mathbb{Z}_3$-symmetric spin chains and $(2+1)$-dimensional $\mathbb{Z}_2$ lattice gauge theory.
Section \ref{sec:conclusion} is then devoted to conclusion and discussion.


\section{Conditions for the ETH-violation by $p$-form symmetry}\label{sec:p-ETH}
In this section, we briefly review how the ETH is broken due to the $p$-form symmetry, and sort out its sufficient conditions in \cite{Fukushima:2023swb}. 
We consider a $(d+1)$-dimensional manifold $\cM\times \mathbb{R}$, where $\cM$ is a $d$-dimensional space manifold.
Let the system have a $G$ $p$-form symmetry with $(d-p)$-dimensional topological symmetry operator, where $G$ is an Abelian group.
Throughout this paper, symmetry operators extend to the spatial directions, and they are represented as unitary operators $U_\alpha(\tilde{C})$ with the support $\tilde{C}\subset\cM$.

\medskip

\begin{figure}[htbp]
\centering
\begin{minipage}[ht]{0.28\linewidth}
\centering
\includegraphics[width=0.8\linewidth]{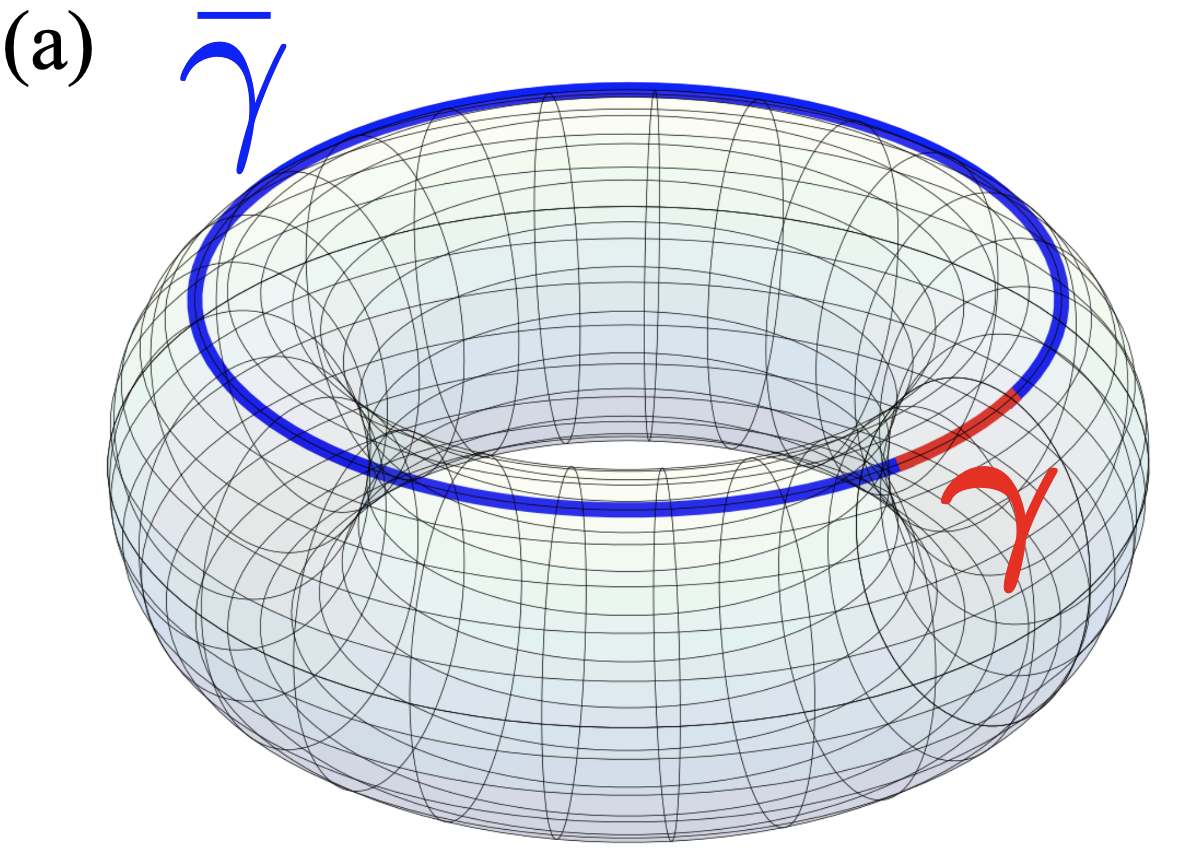}
\end{minipage} 
\hspace{-10pt}
\begin{minipage}[ht]{0.3\linewidth}
\centering
\includegraphics[width=1.1\linewidth]{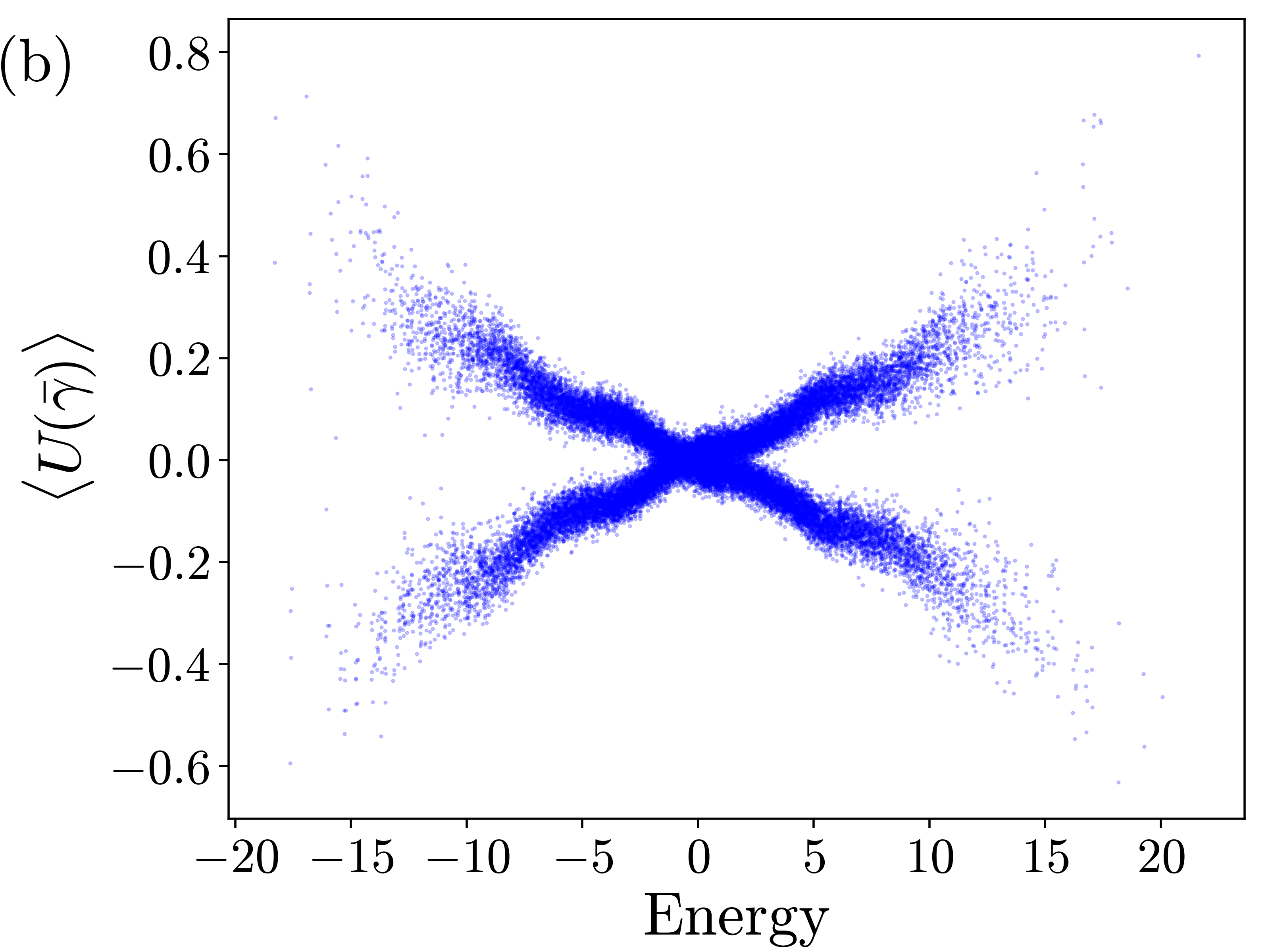}
\end{minipage} 
\hspace{10pt}
\begin{minipage}[ht]{0.3\linewidth}
\centering
\includegraphics[width=1.1\linewidth]{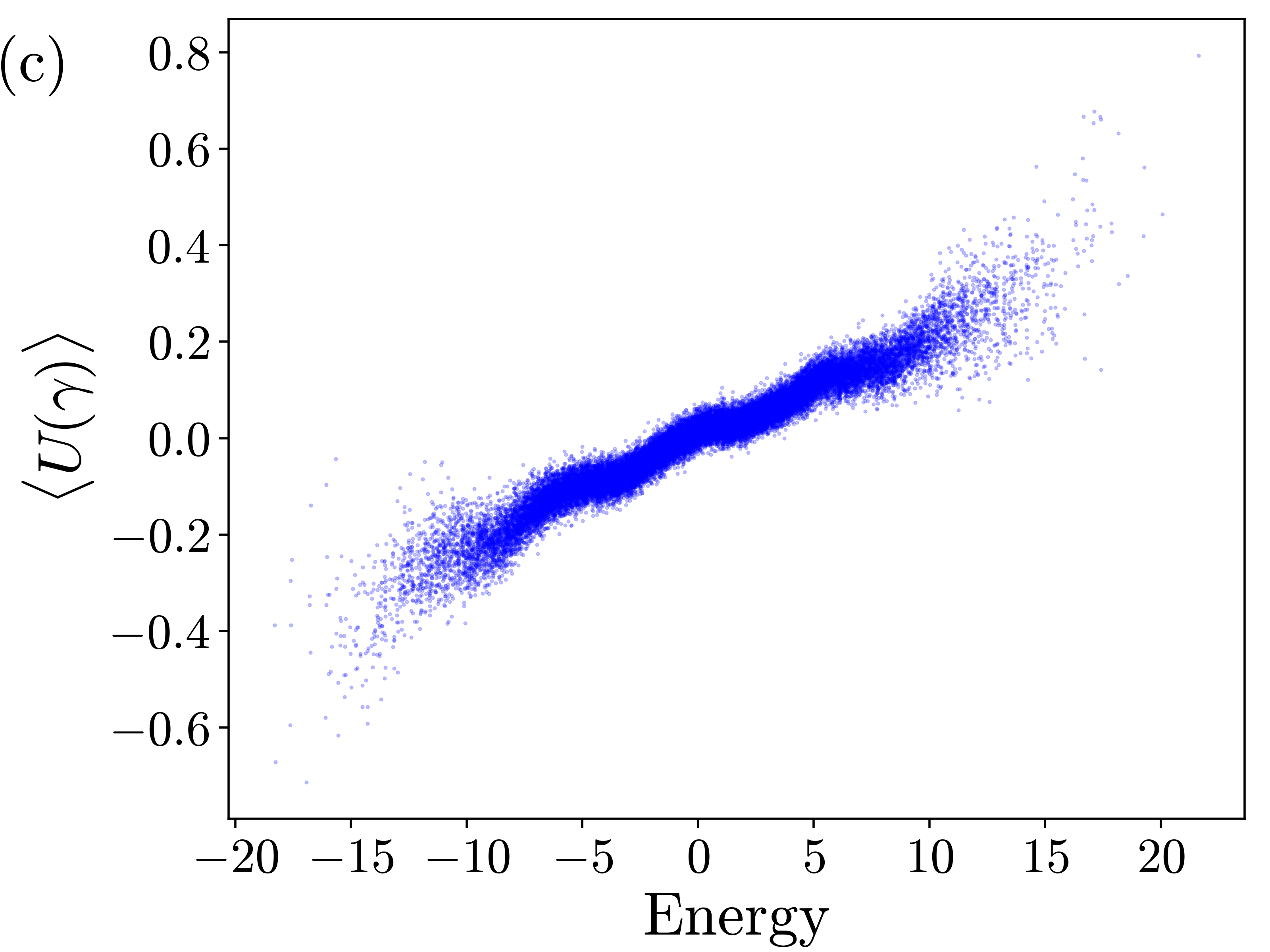}
\end{minipage} 
\caption{\footnotesize (a) Schematics of $\gamma$ and $\bar{\gamma}$ for $\cM=T^2$. The union of $\gamma$ and $\bar{\gamma}$ constitutes a closed manifold $\tilde{C}$.\\
(b)(c) The expectation values of the operator $U(\bar{\gamma})$ and $U(\gamma)$ with respect to the energy eigenstates for the $\mathbb{Z}_2$ gauge theory.
The ETH for $U(\bar{\gamma})$ can be seen violated because there are deviations for a fixed energy $E$, while the ETH for $U(\gamma)$ holds.
}\label{fig:demo}
\end{figure}

The main claim in \cite{Fukushima:2023swb} states that higher-form symmetry of a non-degenerate Hamiltonian leads to the breakdown of the ETH for nontrivial $(d-p)$-dimensional operators.
To show this statement, we assume the following:
\begin{enumerate}
\item[i)]
The symmetry operator $U_\alpha(\tilde{C})$ can be decomposed as $U_\alpha(\tilde{C}) = U_\alpha(\gamma) U_\alpha(\bar{\gamma})$ for an arbitrary $(d-p)$-dimensional submanifold $\gamma\: (\subset \tilde{C})$ and the complement $\bar{\gamma}:= \tilde{C}\backslash \gamma$ (see Fig. \ref{fig:demo} (a)).
This implies the operator with boundaries $U(\gamma)$ and $U(\bar{\gamma})$ are well-defined (not-null) operators.
%
\item[ii)]
An energy shell $[E,E+\delta E]$ contains eigenstates in different symmetry sectors defined by $U_{\alpha}(\tilde{C})$, i.e.,
for at least one nontrivial closed surface, say $\tilde{C}\,(\subset \cM)$, 
there exist energy eigenstates $|E_n\rangle$, $|E_m\rangle$ with $E_n,E_m\in[E,E+\delta E]$ such that $\langle E_n | U_{\alpha}(\tilde{C}) | E_n \rangle\neq \langle E_m | U_{\alpha}(\tilde{C}) | E_m \rangle$.
%
\item[iii)]
Given an energy shell $[E,E+\delta E]$, the microcanonical average $\langle U_{\alpha}(\gamma)\rangle_\mathrm{mc}^{\delta E}(E)$ takes a nonzero value in the thermodynamic limit.
\end{enumerate}
It follows that either $U_\alpha(\gamma)$ or $U_\alpha(\bar{\gamma})$ necessarily breaks the ETH within the energy shell $[E,E+\Delta E]$ under the above condition\footnote{
Although the operators $U_\alpha(\gamma)$ and $U_\alpha(\bar{\gamma})$ are not Hermitian in general,
the ETH-violation for these operators entails ETH-violation for certain Hermitian operators as well.
Indeed, 
the breakdown of the ETH for either of the Hermitian operators $U_\alpha(\bar{\gamma})+U_\alpha^\dag(\bar{\gamma})$ or $i(U_\alpha(\bar{\gamma})-U_\alpha^\dag(\bar{\gamma}))$ follows from the ETH-violation for $U_\alpha(\bar{\gamma})$.
}. 
It can be shown as follows (the proof for more general case can be found in Supplemental Material of \cite{Fukushima:2023swb}). 
We first consider a $(d-p)$-dimensional surface $\gamma$ with boundary, which satisfies the property iii).
In the case where $U_{\alpha}(\gamma)$ does not satisfy the ETH, our claim holds in the first place; we thus suppose $U_{\alpha}(\gamma)$ satisfies the ETH, i.e., 
\begin{align}
\langle E_n | U_{\alpha}(\gamma) | E_n \rangle\simeq \langle E_m | U_{\alpha}(\gamma) | E_m \rangle\simeq \langle U_{\alpha}(\gamma)\rangle_\mathrm{mc}^{\Delta E}(E).
\end{align}
The Hamiltonian $H$ is assumed to have no degeneracy, and thus its eigenstates $|E_n\rangle$, $|E_m\rangle$ are eigenstates of $U_{\alpha}(\tilde{C})$ as well.
Since the group $G$ is abelian,  the eigenvalues are expressed as
\begin{align}
U_{\alpha}(\tilde{C})|E_n\rangle=e^{i\alpha q_n}|E_n\rangle, \qquad
U_{\alpha}(\tilde{C})|E_m\rangle=e^{i\alpha q_m}|E_m\rangle,
\end{align}
where $q_n,q_m\in\mathbb{R}$. 
The assumption ii) now indicates that $|E_n\rangle$ and $|E_m\rangle$ belong to different sectors, i.e., $e^{i\alpha q_n}\neq e^{i\alpha q_m}$. 
The definition of $\bar{\gamma}$ leads to 
\begin{align}
\langle E_n |U_{\alpha}^{-1}(\bar{\gamma}) | E_n\rangle = \langle E_n | U_{\alpha}(\gamma)U_{\alpha}(\tilde{C})^{-1} | E_n\rangle = e^{-i\alpha q_n}\langle E_n |  U_{\alpha}(\gamma)| E_n\rangle
\end{align}
and 
$\langle E_m |U_{\alpha}^{-1}(\bar{\gamma}) | E_m\rangle = e^{-i\alpha q_m}\langle E_m | U_{\alpha}(\gamma)| E_m\rangle$.
Recalling iii) and the supposition of the ETH, i.e., 
\begin{align}
\langle E_n |  U_{\alpha}(\gamma) | E_n \rangle\simeq \langle E_m |  U_{\alpha}(\gamma) | E_m \rangle\simeq \langle U_{\alpha}(\gamma)\rangle_\mathrm{mc}^{\Delta E}(E)\neq 0,
\end{align}
we obtain the relation
\begin{align}
\langle E_n |U_{\alpha}^{-1}(\bar{\gamma}) | E_n\rangle \neq \langle E_m |  U_{\alpha}^{-1}(\bar{\gamma}) | E_m\rangle
\quad \Rightarrow \quad 
\langle E_n | U_{\alpha}(\bar{\gamma})  | E_n\rangle \neq \langle E_m | U_{\alpha}(\bar{\gamma}) | E_m\rangle.
\label{ETH-violation}
\end{align}
After all, we see that $U_{\alpha}(\bar{\gamma}) $ violates the ETH, and the claim has been proven.

\medskip

Note here that symmetries do not automatically lead to the degeneracy of the spectrum and the mixing of the symmetry sectors. As a simple example, we can consider a Hamiltonian $H=\operatorname{diag}(1,1,-1,-2)$ and a charge $Q=\operatorname{diag}(1,1,1,-1)$. The operators $H$ and $Q$ commute each other, but there are energy eigenstates without degeneracy, and the eigenstates with $H=1$ do not exhibit the mixing of the symmetry sectors. 
However, in the presence of symmetries with nontrivial projective phases, we can always obtain the degeneracies and the mixture of the symmetry sectors as discussed in Section \ref{sec:Projective}, and this is a key to the following discussion.

\medskip

Let us comment on the volume of the ``bath'' for the ETH-breaking observables (say, $U_\alpha(\bar{\gamma})$).
Let $V_{\bar{\gamma}}$ and $V_{\cM\backslash\bar{\gamma}}$ denote the volume of $\bar{\gamma}$ and $\cM\backslash\bar{\gamma}$, respectively.
For the operator $U_\alpha(\bar{\gamma})$ with the $(d-p)$-dimensional support, the $d$-dimensional complements $\cM\backslash\bar{\gamma}$ can be regarded as the bath.
If we consider a $0$-form symmetry, the ratio $V_{\bar{\gamma}}/V_{\cM\backslash\bar{\gamma}}$ remains finite in the thermodynamic limit $V:=V_{\bar{\gamma}}\cup V_{\cM\backslash\bar{\gamma}}\to\infty$, and thus the breakdown of the ETH may be attributed to the smallness of the bath.
%
In contrast, for higher-form symmetry with $p\geq 1$, the volumes scales as $V_{\bar{\gamma}}/V_{\cM\backslash\bar{\gamma}}\rightarrow 0$ in the thermodynamic limit
since $\bar{\gamma}$ and $\cM\backslash\bar{\gamma}$ are $(d-p)$-dimensional and $d$-dimensional, respectively.
Thus, the higher-form symmetry hinders thermalization even when the support of the observable is much smaller than its bath.

\medskip

Figure \ref{fig:demo} (b) shows the numerical result for the $(2+1)$-dimensional $\mathbb{Z}_2$ gauge theory, which is described in detail in Section \ref{sec:Z2 gauge}.
The 1-dimensional operator $U(\bar{\gamma}):=U_1(\bar{\gamma})$ violates the ETH while the ETH for the local operator $U(\gamma)$ holds.
Although the conditions i), ii) and iii) are indeed satisfied in this case, it is generally challenging to confirm whether the conditions, especially ii), are satisfied without explicit numerical calculation.
In the following sections, we show that the condition ii) is satisfied when the system incorporate a projective representation for the symmetry of interest and an auxiliary symmetry that is to be broken by perturbations for the Hamiltonian.

\medskip

We comment on a related framework referred to as the subsystem eigenstate thermalization hypothesis (ETH)\cite{Dymarsky:2016ntg}. 
The subsystem ETH claims energy eigenstates can be regarded as the microscopic thermal equilibrium (MITE)\cite{goldstein2015thermal,goldstein2017macroscopic}, where the reduced density matrix tends to the microcanonical density matrix in the thermodynamic limit. This condition is stronger than the ETH for operators with a given small support. Thus, the violation of the ETH for $U(\bar{\gamma})$ also indicates the breakdown of the subsystem ETH with respect to the conventional microcanonical ensemble.


\section{Projective representation for Abelian group}\label{sec:Projective}

In this section, we show that the condition ii) can be always satisfied if a $\mathbb{Z}_N\times\mathbb{Z}_N$ symmetry with a projective representation is explicitly broken by perturbations.
Before delving into such perturbations, we first discuss the degeneracy due to projective representations.

\subsection{Degeneracy by projective representation}

Let $G$ be an Abelian group and the theory have a $G$ symmetry.
In the operator formalism, a projective representation of the $G$ symmetry is realized as
\begin{align}\begin{split}
    U_{g_1}U_{g_2}= e^{i\phi(g_1,g_2)} U_{g_1g_2}, \qquad g_1,g_2\in G.\\
    U_{g_1}U_{g_2} = e^{i\phi(g_1,g_2) - i\phi(g_2,g_1)} U_{g_2}U_{g_1},
\end{split}\end{align}
where $U_{g_1}$, $U_{g_2}$ are unitary operators, and $\phi:G\times G\to \mathbb{R}$ is the projective phase.
Since the theory has the $G$ symmetry, the Hamiltonian $H$ commutes with the unitary operators $[H,U_g]=0,\;{}^{\forall}g\in G$.
The non-vanishing projective phase with $\exp(i(\phi(g_1,g_2) - \phi(g_2,g_1)))\neq 1$ immediately leads to the degeneracy of arbitrary eigenstates of the Hamiltonian. This is because if you have a simultaneous eigenstate s.t. $H\ket{E}=E\ket{E}$ and $U_{g_1}\ket{E}=e^{i\alpha}\ket{E}, \alpha\in\mathbb{R}$, we obtain
\begin{align}
    \bra{E}U_{g_2}\ket{E}=\bra{E}U_{g_1}^\dagger U_{g_2}U_{g_1}\ket{E} 
    =e^{-(i\phi(g_1,g_2) - i\phi(g_2,g_1))}\bra{E}U_{g_2}\ket{E}
    \quad \Rightarrow \quad
    \bra{E}U_{g_2}\ket{E}=0.
\end{align}
Since $\ket{E}$ and $U_{g_2}\ket{E}$ are orthogonal to each other, they are degenerate energy eigenstates with the eigenvalue $E$.
Note that $\ket{E}$ and $U_{g_2}\ket{E}$ belong to different symmetry sector of $U_{g_1}$:
\begin{align}
U_{g_1}( U_{g_2} \ket{E} ) = e^{i\phi(g_1,g_2) - i\phi(g_2,g_1)} e^{i\alpha}(U_{g_2}\ket{E}),
\end{align}
where $\alpha$ is the charge for the state $\ket{E}$.

\medskip

We define simultaneous eigenstates of the Hamiltonian $H$ and $U_{g_1}$ by
\begin{align}
    & H \ket{E,\alpha} = E \ket{E,\alpha}, \quad H \ket{E,\beta} = E \ket{E,\beta},\\
    & U_{g_1}\ket{E,\alpha} = e^{i\alpha}\ket{E,\alpha}, \quad U_{g_1} \ket{E,\beta} = e^{i\beta}\ket{E,\beta},
\end{align}
The charged operator under the symmetry $G$ can be also introduce as
\begin{align}
U_{g_1}^\dagger W_q U_{g_1} = e^{iq_{g_1}}W_q, \quad 
U_{g_2}^\dagger W_q U_{g_2} = e^{iq_{g_2}}W_q.
\end{align}



\medskip

Now we consider the matrix elements of the charged operator $W_q$.
The diagonal part satisfies
\begin{align}
    &\bra{E,\alpha} W_q \ket{E,\alpha} = 
    \bra{E,\alpha} U_{g_1}^\dagger W_q U_{g_1} \ket{E,\alpha}e^{-iq_{g_1}}
    = \bra{E,\alpha} W_q \ket{E,\alpha}e^{-iq_{g_1}},\no\\
    \Rightarrow\quad &
    \bra{E,\alpha} W_q \ket{E,\alpha} = 0,
\end{align}
if the operator $W_q$ is nontrivially charged under the action of $g_1$, i.e., $e^{-iq_{g_1}}\neq 1$.
On the other hand, the operator with trivial charge under $g_1$, but charged under $g_2$, i.e., $e^{-iq_{g_2}}\neq 1$ can have the nonvanishing expectation value in this basis while the off-diagonal part with $\alpha\neq\beta$ necessarily vanishes:
\begin{align}
    &\bra{E,\alpha} W_q \ket{E,\beta} = \bra{E,\alpha}U_{g_1}^\dagger W_q U_{g_1} \ket{E,\beta}
    =e^{i(\beta-\alpha)}\bra{E,\alpha} W_q \ket{E,\beta},\no\\
    \Rightarrow\quad& \bra{E,\alpha} W_q \ket{E,\beta} = 0.
    \label{off-diag}
\end{align}
We stress that all of the properties discussed here can be applied not only to the ground states but also to arbitrary energy eigenstate, although topological robustness of degeneracy does not hold for general eigenstates since the gaps are exponentially small.

\subsection{Symmetry violating perturbation}
We now discuss a consequence of weak breaking of symmetries with a projective representation. To this end, let $G_1$ and $G_2$ be Abelian groups and the group $G=G_1\times G_2$ projectively acts on the Hilbert space of the theory. 
The corresponding unitary operator is given by $U_{g_1}$ and $\tilde{U}_{g_2}$ for $G_1$ and $G_2$, respectively.
We consider a situation such that each of the symmetry is realized by a standard linear representation, but they have nontrivial projective phases between them: $U_{g_1}\tilde{U}_{g_2}=e^{i\phi(g_1,g_2)}\tilde{U}_{g_2}U_{g_1}$.
We perturb the Hamiltonian by adding a term $\sum_{\rm site}W_{q}$, where $W_q$ is a charged operator under $G_2$ but trivially transforms under $G_1$, i.e.,
\begin{align}
    U_{g_1}^\dagger W_q U_{g_1} = W_q,\quad
    \tilde{U}_{g_2}^\dagger W_q \tilde{U}_{g_2} = e^{iq_{g_2}}W_q,\qquad
    g_1\in G_1,\; g_2\in G_2.
\end{align}
Here, we assume the the operator $W_q$ is a local operator.

\medskip

In the following, we focus on the system with a discrete spectrum realized by appropriate regularizations.
For brevity, we specify $G_1 = G_2 = \mathbb{Z}_N$ as the symmetry groups and thus the charge of $W_q$ is simply given by $q_{m}=qm$ for $m\in G_2=\mathbb{Z}_N$.
As explained in the previous subsection, arbitrary energy eigenstates are degenerate, and then we work with the simultaneous eigenbasis of the unperturbed Hamiltonian $H$ and $U_{g_1}$ (${}^{\forall}g_1\in G_1=\mathbb{Z}_N$), i.e., $H\ket{E,\alpha} = E\ket{E,\alpha}$, $U_{g_1}\ket{E,\alpha}= e^{i\alpha}\ket{E,\alpha}$.
Note that the number of the degeneracy of each energy eigenstate is at least $N$ as long as $\exp(i(\phi(g_1,g_2) - \phi(g_2,g_1)))\neq 1$ with ${}^{\forall}g_1\in G_1$, ${}^{\forall}g_2\in G_2$, since 
$\tilde{U}_m\ket{E,\alpha}$ ($m\in G_2=\mathbb{Z}_N$) are eigenstates of $U_{g_1}$ with different eigenvalues, and $\bra{E,\alpha}\tilde{U}_{m}^{\dagger}\tilde{U}_n\ket{E,\alpha}=\bra{E,\alpha}\tilde{U}_{n-m}\ket{E,\alpha}=0$ for $n\neq m$.
Although extra degeneracies are possible in general, such accidental degeneracies can be removed by deforming the Hamiltonian while preserving the $\mathbb{Z}_N\times\mathbb{Z}_N$ symmetry.
We thus assume that all of the energy eigenstates are $N$-fold degenerate in the following discussion.
The degenerate subspace $\cH(E):=\operatorname{span}\{\ket{E,\alpha}\,|\,\alpha=0,1,\dots,N-1\}$ is also expressed as $\cH(E)=\operatorname{span}\{\tilde{U}_{g_2}\ket{E,\alpha},{}^{\forall}g_2\in G_2\}$.

\medskip

 The perturbed Hamiltonian is defined by
\begin{align}
    \tilde{H}(\lambda):= H+ \lambda H_1, \qquad 
    H_1 : =  \sum_{j:\;\mathrm{site}}\frac{W_q(j)+W_q(j)^\dagger}2\,,
\end{align}
where $\lambda$ is the perturbation parameter.
After this perturbation, the system with $\tilde{H}(\lambda)$ only exhibits $G_1$ symmetry since the operator $W_q(j)$ has a trivial charge under $G_1$.
The perturbation part $H_1$ is diagonalized by $\ket{E,\alpha}$ basis in the subspace $\cH(E)$ since the off-diagonal part $\bra{E,\alpha}W_q\ket{E,\beta}$ ($\alpha\neq\beta$) always vanishes as in (\ref{off-diag}). 
In order to estimate the energy modification, we utilize the Hellmann-Feynman theorem for degenerate spectra \cite{zhang2002breakdown,vatsya2004comment,fernandez2004comment,balawender2004comment, zhang2004extended,fernandez2019hellmann}.
Once the operator $d\tilde{H}(\lambda)/d\lambda= H_1$ is diagonalized in the subspace $\cH(E)$, we can obtain
\begin{align}
    \frac{d E(\alpha;\lambda)}{d\lambda} = \bra{E,\alpha;\lambda}H_1 \ket{E,\alpha;\lambda}
    =
    \sum_{j:\;\mathrm{site}}\Re \bra{E,\alpha;\lambda}W_q(j) \ket{E,\alpha;\lambda},
    \label{HFT}
\end{align}
where $E(\alpha;\lambda)$ is the energy eigenvalue for the eigenstate that depends on the parameter $\lambda$: $E(\alpha;\lambda)\ket{E,\alpha;\lambda}=\tilde{H}(\lambda)\ket{E,\alpha;\lambda}$.
In the first order, the perturbed energy reads
\begin{align}
    E(\alpha;\lambda)
    = E + \lambda\sum_{j:\;\mathrm{ site}}\Re\bra{E,\alpha}W_q(j)\ket{E,\alpha} + \cO(\lambda^2).
    \label{perturbed energy}
\end{align}
Significantly, for all elements of $\cH(E)$, the expectation $\bra{E,\alpha;\lambda}W_q\ket{E,\alpha;\lambda}$ have different values because of the relation $\bra{E,\alpha;\lambda}\tilde{U}_{m}^\dagger W_q \tilde{U}_{m}\ket{E,\alpha;\lambda} = e^{iqm}\bra{E,\alpha;\lambda}W_q\ket{E,\alpha;\lambda}$.
Except for the case $\bra{E,\alpha;\lambda}\tilde{U}_m^\dagger W_q \tilde{U}_m\ket{E,\alpha;\lambda} = {\bra{E,\alpha;\lambda}W_q\ket{E,\alpha;\lambda}}^*$, we can see that the perturbed energies (\ref{perturbed energy}) are split for $\ket{E,\alpha;\lambda}$ and $\tilde{U}_m\ket{E,\alpha;\lambda}$.
Even if the energies are still degenerate in the first order perturbations, the degeneracies are lifted by higher order perturbations due to the mixing with other energy eigenstates.

\medskip

Note here that the standard perturbation theory for higher order breaks down in the large system-size limit $V\to\infty$. Since the separations of the energies tend to be exponentially small in the limit, they become small enough compared to the perturbation, i.e., $E-E'\simeq \lambda \bra{E,\alpha}H_1\ket{E,\alpha}$, where $E'$ is an energy eigenvalues of another eigenstate.
The higher-order perturbation is no longer valid unless $\lambda \bra{E,\alpha}H_1\ket{E,\alpha}\simeq \cO(e^{-V})$, and thus we have to resort to the Hellmann-Feynmann theorem as in (\ref{HFT}).

\medskip

\begin{figure}[htbp]
\centering
\includegraphics[width=0.8\linewidth]{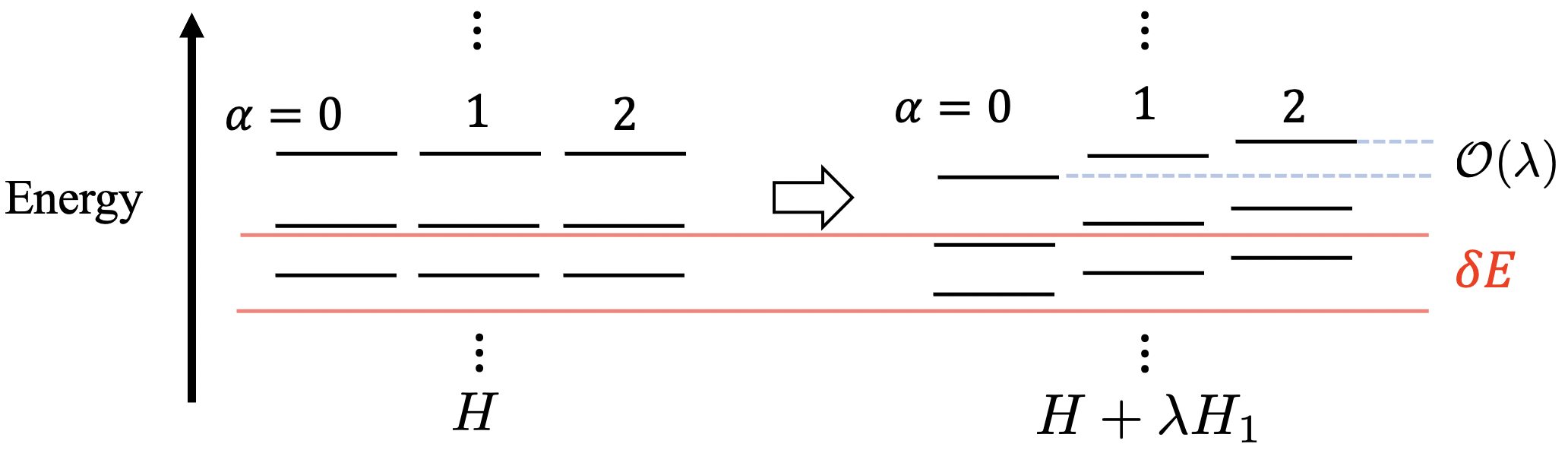}
\caption{\footnotesize Schematics for the spectrum.
$\alpha$ denotes the charge for $U_m$ s.t. $U_m\ket{E,\alpha}=e^{i\alpha}\ket{E,\alpha}$.
After the perturbation with sufficiently small $\lambda$, the degeneracies are resolved so that the condition ii) is satisfied.
}\label{fig:split}
\end{figure}

After all, we obtain the energy eigenstates with the energy separations of order $\lambda$ (see Fig.~\ref{fig:split}).
The key point is that these eigenstates in $\cH(E)$ have distinct charges of $U_{m}$.
In the thermodynamics limit $V\to \infty$, we should take the width of the energy shell $\delta E=\cO(V^{1/2})$, and we suppose the energy splitting also scales as $\cO(V^{1/2})$, i.e.,
\begin{align}
    \lambda\left(\sum_{j:\;\mathrm{ site}}\Re\bra{E,\alpha}W_q(j)\ket{E,\alpha}
    -\sum_{j:\;\mathrm{ site}}\Re\bra{E,\beta}W_q(j)\ket{E,\beta}\right)
    \simeq \cO(V^{1/2}),
    \qquad
    \alpha\neq\beta.
    \label{scale-supposition}
\end{align}
This relation is expected to be naturally realized, but we can also force (\ref{scale-supposition}) by e.g., setting $\lambda\simeq\cO(V^{-1/2})$ since $W_q$ is a local operator. 
Another approach is to implement a weak randomness such that $\lambda H_1=\lambda \sum_j r_j W_q(j)$, where $r_j$ is a uniformly chosen constants from $[-r,r]$, $0<r\ll 1$.
Assuming the expecation value $\langle W_q(j)\rangle$ is almost uniform, we obtain the variation $\langle H_1 \rangle\simeq \cO(V^{1/2})$.
Under the supposition (\ref{scale-supposition}),
we notice that given an energy window with the width $\delta E$, we can arrange eigenstates with different symmetry sectors with respect to $G_1$ within it by tuning the parameter $\lambda$.
This exactly indicates the condition ii) is satisfied.

\section{Demonstration for lattice models}\label{sec:demonstration}
In this section, we demonstrate the statement discussed in Section \ref{sec:Projective} by numerically calculating the energy spectra for concrete examples.

\subsection{$(1+1)$-dimensional spin chains}
As the first example, we consider a $(1+1)$-dimensional $\mathbb{Z}_2\times\mathbb{Z}_2$-symmetric spin chain and its $\mathbb{Z}_3\times\mathbb{Z}_3$-symmetric generalization.
Both of the models only exhibit 0-form symmetries, and thus the ETH-violating operators based on the mechanism in \cite{Fukushima:2023swb} are $1$-dimensional.
Even though one can not tell whether ETH violation is caused by the smallness of the baths for such operators that have the same dimensionality with the space, it is instructive to illustrate that the discussion in Section \ref{sec:Projective} indeed holds for those models.

A projective representation of $\mathbb{Z}_N\times\mathbb{Z}_N$ is realized on the $N$-dimensional Hilbert space spanned by $\ket{g}$, $g=0,1,\dots,N-1$. 
The generators of each $\mathbb{Z}_N$ are represented by ``clock'' operators $Z$ and ``shift'' operators $X$, which satisfy the relations \cite{Alavirad:2019iea}
\begin{align}
 ZX=e^{\frac{2\pi i}{N}}XZ.
\end{align}
The operators act on the Hilbert space as
\begin{align}
    Z\ket{g}=e^{2\pi i\frac{g}{N}}\ket{g},\qquad
    X\ket{g}=\ket{g+1\!\!\!\mod N}.
\end{align}
In the matrix form, they can be explicitly expressed as
\begin{align}
    Z=\left(\begin{array}{ccccc}
    1 & 0 & 0 & \cdots & 0 \\
    0 & e^{2\pi i\frac{1}{N}} & 0 & \cdots & 0 \\
    0 & 0 &  e^{2\pi i\frac{2}{N}} & \cdots & 0 \\
    \vdots & \vdots & \vdots & \ddots & \vdots \\
    0 & 0 & 0 & \cdots & e^{2\pi i\frac{N-1}{N}}
    \end{array}\right),
    \qquad
    X=\left(\begin{array}{ccccc}
    0 & 0 & \cdots & 0 & 1 \\
    1 & 0 & \cdots & 0 & 0 \\
    0 & 1 & \cdots & 0 & 0 \\
    \vdots & \vdots &\ddots & \vdots & \vdots\\
    0 & 0 & \cdots & 1 & 0
    \end{array}\right).
\end{align}
For $N=2$, these are just the standard Pauli matrices. 

\medskip 

In order to accommodate a spin chain with $L$ sites $(L>N)$, we introduce the tensor product $\ket{g}\otimes\dots\otimes \ket{g}$, and the operators acting only the $j$-th site $Z_j:=\bm{1}\otimes\dots\otimes Z\otimes\dots\otimes\bm{1}$ and $X_j:=\bm{1}\otimes\dots\otimes X\otimes\dots\otimes\bm{1}$.
The symmetry operators are then given by 
\begin{align}
    U_1:=\prod_{j=1}^L Z_j,
    \qquad
    \tilde{U}_1:=\prod_{j=1}^L X_j.
\end{align}
We note that the projective phase between $U_i$ and $\tilde{U}_j$ can be trivial for the case $\gcd(N,L)=N$ ($\Leftrightarrow N|L$), and thus we suppose $\gcd(N,L)\neq N$ so that the discussion in the previous section always holds.
This projective phase can be interpreted as the $\mathbb{Z}_N$ action by $U_i$ on the other symmetry operators $\tilde{U}_i$, i.e.,
$U_j \tilde{U}_k U_j^{-1} \exp(2\pi i\cdot jk/N) U_k$.
Considering a central extension of $\mathbb{Z}_N\times\mathbb{Z}_N$ by $\mathbb{Z}_{\operatorname{lcm}(L,N)/L}$, we can realize this relation as a linear representation, where the center $\mathbb{Z}_{\operatorname{lcm}(L,N)/L}$ is generated by
\begin{align}
    \prod_{j=1}^L e^{\frac{2\pi i}{N}} = e^{\frac{2\pi iL}{N}}.
\end{align}

\subsubsection*{$\mathbb{Z}_2\times\mathbb{Z}_2$-symmetric spin chain}
In the $N=2$ case, a $\mathbb{Z}_2\times\mathbb{Z}_2$-invariant $(1+1)$-dimensional spin chain is given by
\begin{align}
    H_{N=2} =  \sum_{j=1}^{L} \left(
    J^x_j X_jX_{j+1} + J^y_j Y_jY_{j+1} + J^z_j Z_jZ_{j+1}
    \right)\,,
\end{align}
where $Y_j:=iX_jZ_j$. 
This Hamiltonian is nothing but the one of the XYZ Heisenberg spin chain.
To remove unwanted spacetime symmetry, we introduce a weak randomness to the couplings $J^x_j$, $J^y_j$ and $J^z_j$.
There is still  an extra symmetry that flips the sign of one of the Pauli matrices, e.g., $Y_j\mapsto -Y_j,$ ${}^\forall j$.
We thus work with a deformed Hamiltonian 
\begin{align}
    H_{N=2} = \sum_{j=1}^{L} \left(
    J^x_j X_jX_{j+1} + J^y_j Y_jY_{j+1} + J^z_j Z_jZ_{j+1}
    \right)
    +
    \alpha\sum_{j=1}^{L} X_jY_{j+1}Z_{j+2}\,,
    \label{2-Ham}
\end{align}
which is indeed $\mathbb{Z}_2\times\mathbb{Z}_2$-symmetric.
Here we consider the periodic boundary condition with $j\sim j+L$, which indicates the topology of the space is $S^1$.

\medskip

\begin{figure}[htbp]
\centering
\begin{minipage}[ht]{0.3\linewidth}
\centering
\includegraphics[width=1\linewidth]{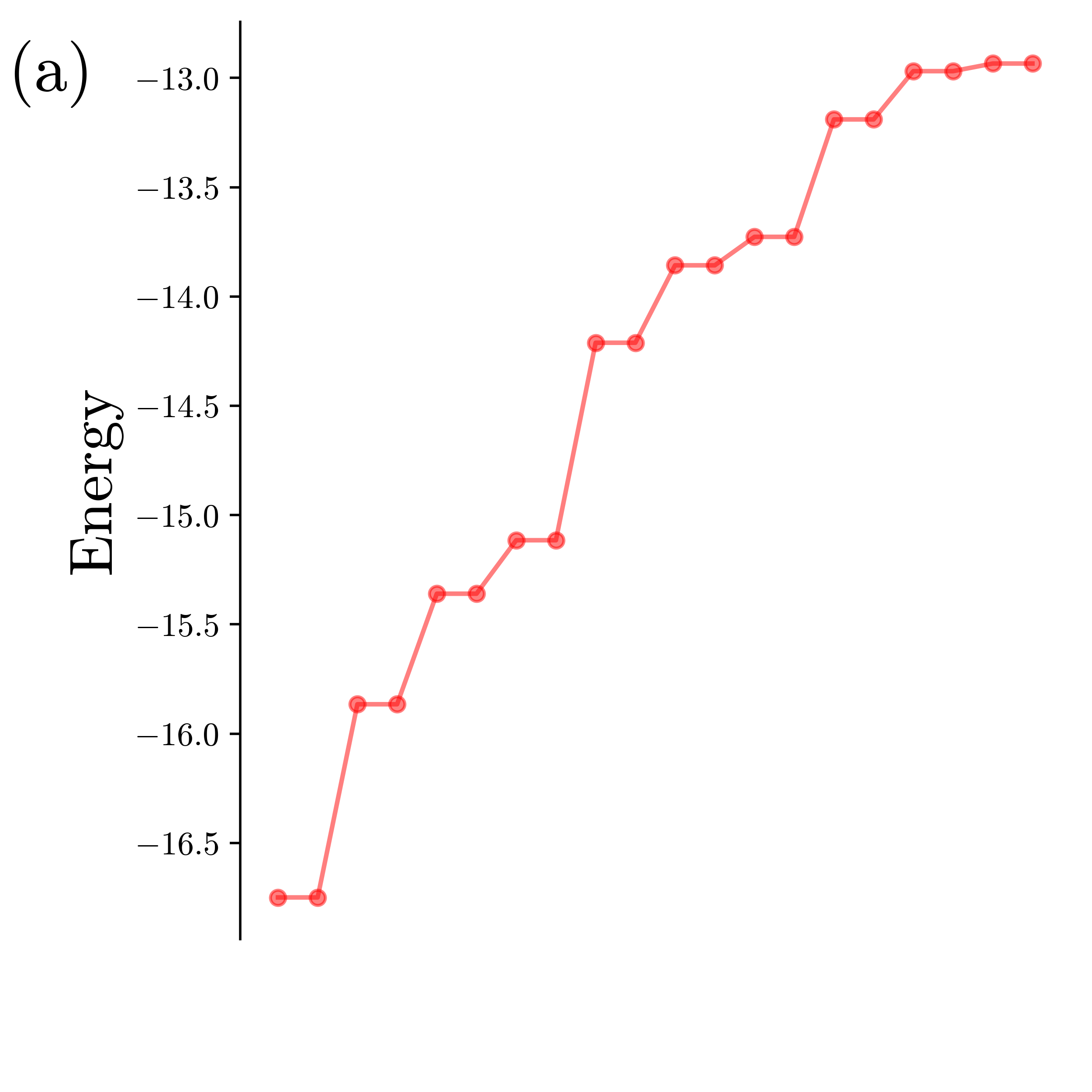}
\end{minipage} 
\begin{minipage}[ht]{0.3\linewidth}
\centering
\includegraphics[width=1\linewidth]{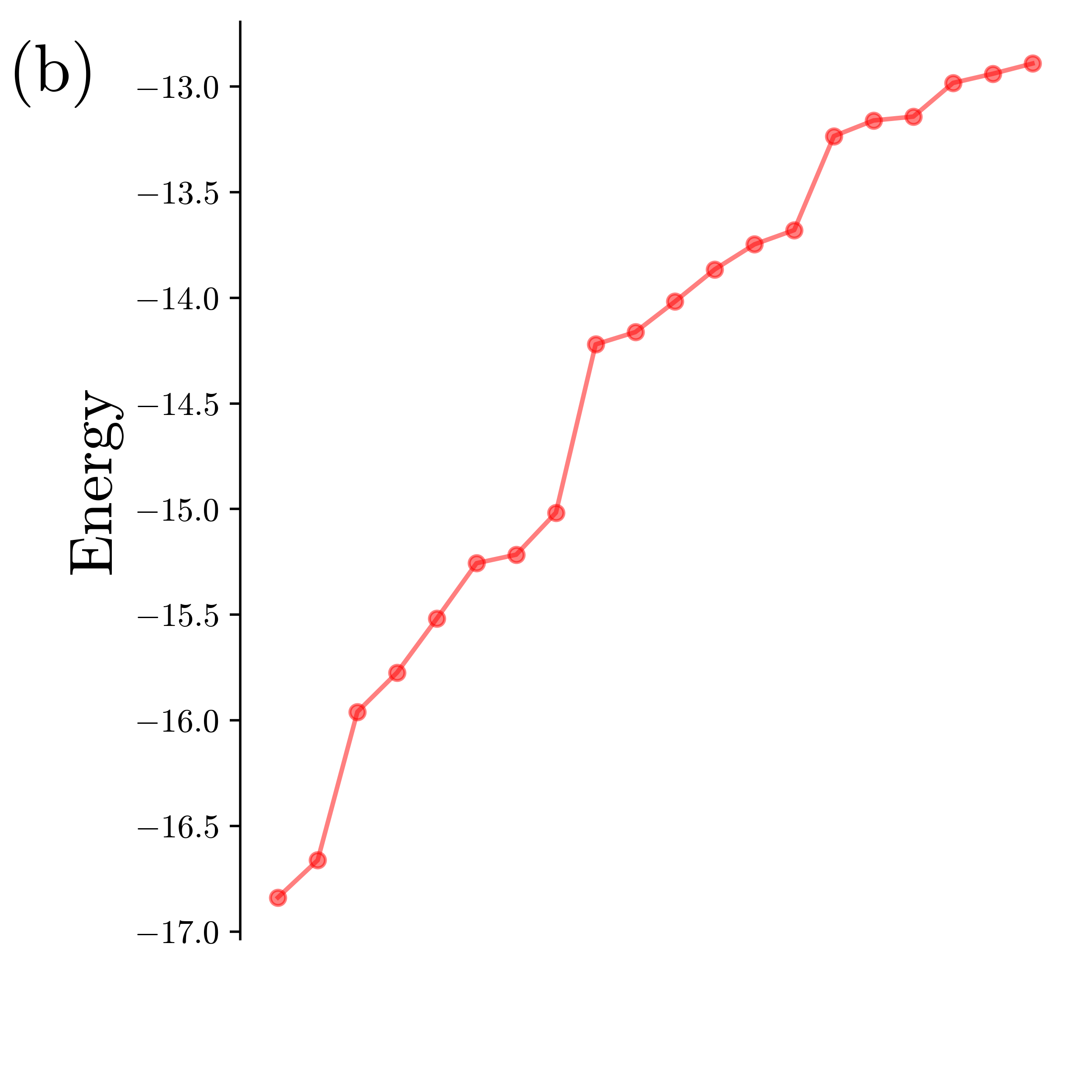}
\end{minipage} 
\begin{minipage}[ht]{0.38\linewidth}
\centering
\includegraphics[width=1\linewidth]{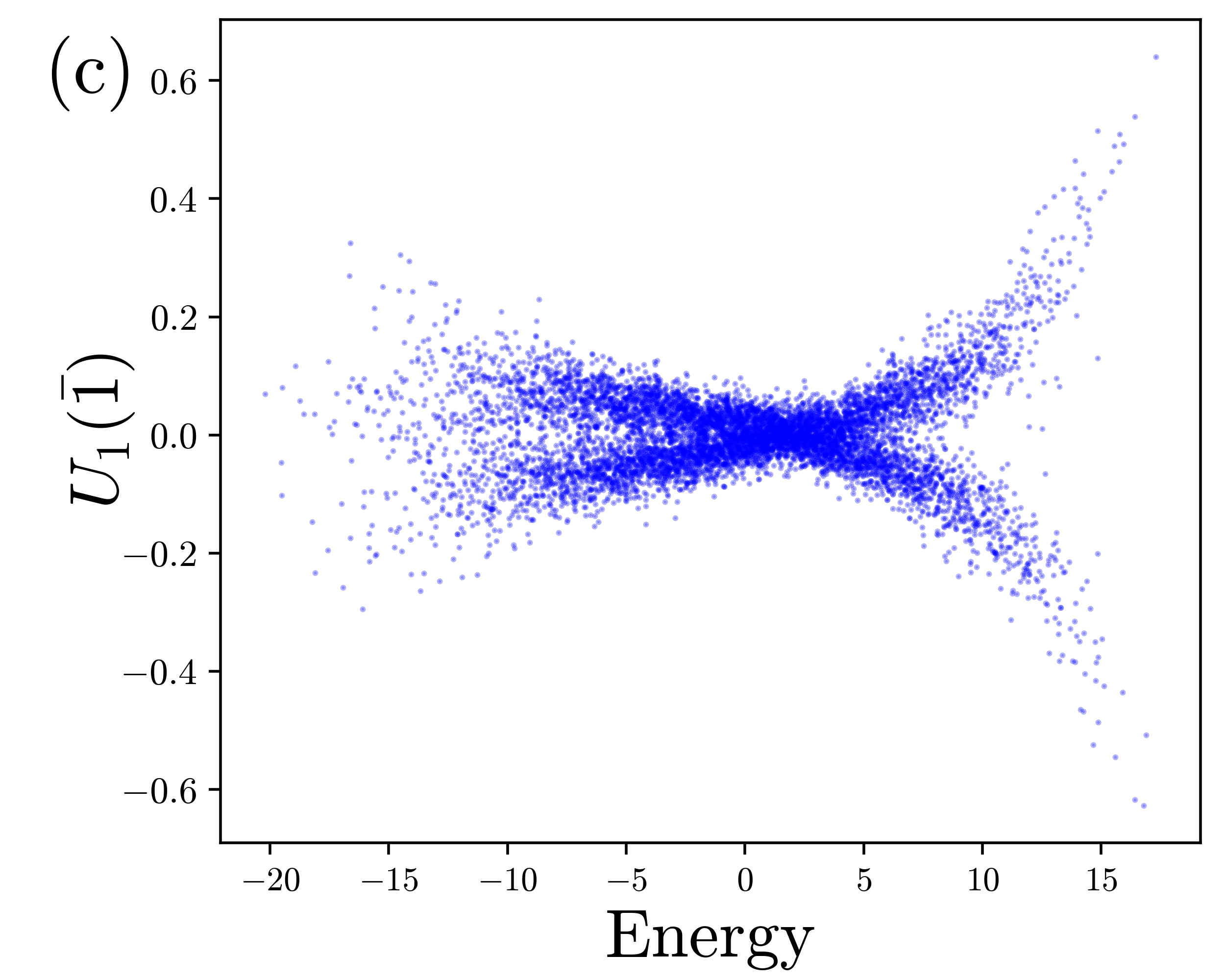}
\end{minipage} 
\caption{\footnotesize (a)(b) Part of energy spectra for $\mathbb{Z}_2\times \mathbb{Z}_2$-symmetry spin chain for $L=11$, $\lambda=0.1$. 
The coupling constants are uniformly distributed in $J_j^x\in[0.9,1.0]$, $J_j^y\in[0.7,0.8]$, $J_j^z\in[0.6,0.7]$, and the parameter is given by $\alpha=0.9$.
The degeneracy in the original Hamiltonian (\ref{2-Ham}) (a) is resolved by the perturbation (\ref{2-Ham-1}) (b).
(c) The expectation value of $U_1(\bar{1})$ for $L=13$, $\lambda=0.4$.
The expectations are separated into two sectors, and thus the ETH for $U_1(\bar{1})$ is not satisfied.
}\label{fig:spin_n2}
\end{figure}

To weakly break the $\mathbb{Z}_2$ generated by $\tilde{U}_1$, we perturb the Hamiltonian as
\begin{align}
    \tilde{H}_{N=2} := H_{N=2} + \lambda \sum_{j=1}^L Z_j\,.
    \label{2-Ham-1}
\end{align}
The numerical results are shown in Fig. \ref{fig:spin_n2}.
Since the surviving symmetry operator is given by $U_1:=\prod_{j=1}^{L}Z_j$, one of the ETH-violating operator in this case is given by $U_1(\bar{1}):=\prod_{j=2}^{L}Z_j$.
After the perturbation, the double degeneracy is completely broken, and these eigenstates leads to mixed symmetry sector in a energy shell.

\subsubsection*{$\mathbb{Z}_3\times\mathbb{Z}_3$-symmetric spin chain}
In the $N=3$ case, the Hamiltonian for a $\mathbb{Z}_3\times\mathbb{Z}_3$-symmetric spin chain is given by
\begin{align}
    H_{N=3}:= \sum_{j=1}^L
    \left(
    J^w_jW_jW_{j+1}^\dagger + J^x_jX_jX_{j+1}^\dagger + J^y_jY_jY_{j+1}^\dagger + J^z_jZ_jZ_{j+1}^\dagger
    \right)
    + (\mbox{h.c.}),
    \label{3-Ham}
\end{align}
where $W_j:=Z^\dagger_j X_j$ and $Y_j:=Z_jX_j$. Again, we take the periodic boundary condition $j\sim j+L.$
If the couplings $J^w_j$, $J^w_j$, $J^w_j$ and $J^w_j$ are weakly random and not real, the theory has no relevant symmetries other than $\mathbb{Z}_3\times\mathbb{Z}_3$ symmetry represented by $U_m$ and $\tilde{U}_m$ ($m=1,2$).
Under these symmetry action, the local operators transform as
\begin{align}\begin{split}
    &U_m^\dagger W_j U_m = e^{\frac{4}{3}\pi mi}W_j,\quad
    U_m^\dagger X_j U_m = e^{\frac{4}{3}\pi mi}X_j,\quad
    U_m^\dagger Y_j U_m = e^{\frac{4}{3}\pi mi}Y_j,\quad
    U_m^\dagger Z_j U_m = Z_j,
    \\
    &\tilde{U}_m^\dagger W_j \tilde{U}_m = e^{\frac{4}{3}\pi mi}W_j,\quad
    \tilde{U}_m^\dagger X_j \tilde{U}_m = X_j,\quad
    \tilde{U}_m^\dagger Y_j \tilde{U}_m = e^{\frac{2}{3}\pi mi}Y_j,\quad
    \tilde{U}_m^\dagger Z_j \tilde{U}_m = e^{\frac{2}{3}\pi mi}Z_j,\quad
\end{split}\end{align}

\medskip

\begin{figure}[htbp]
\centering
\begin{minipage}[ht]{0.3\linewidth}
\centering
\includegraphics[width=1\linewidth]{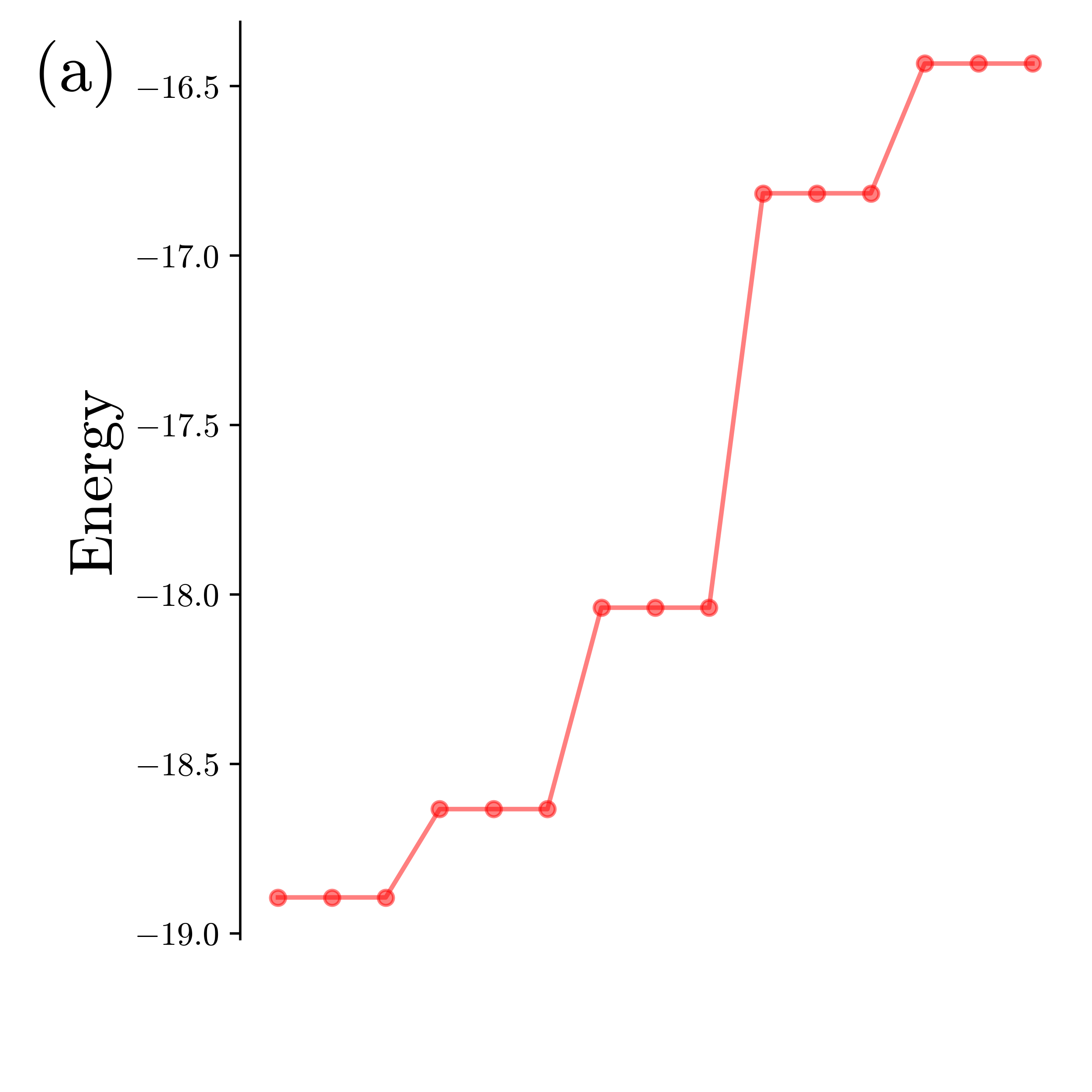}
\end{minipage} 
\begin{minipage}[ht]{0.3\linewidth}
\centering
\includegraphics[width=1\linewidth]{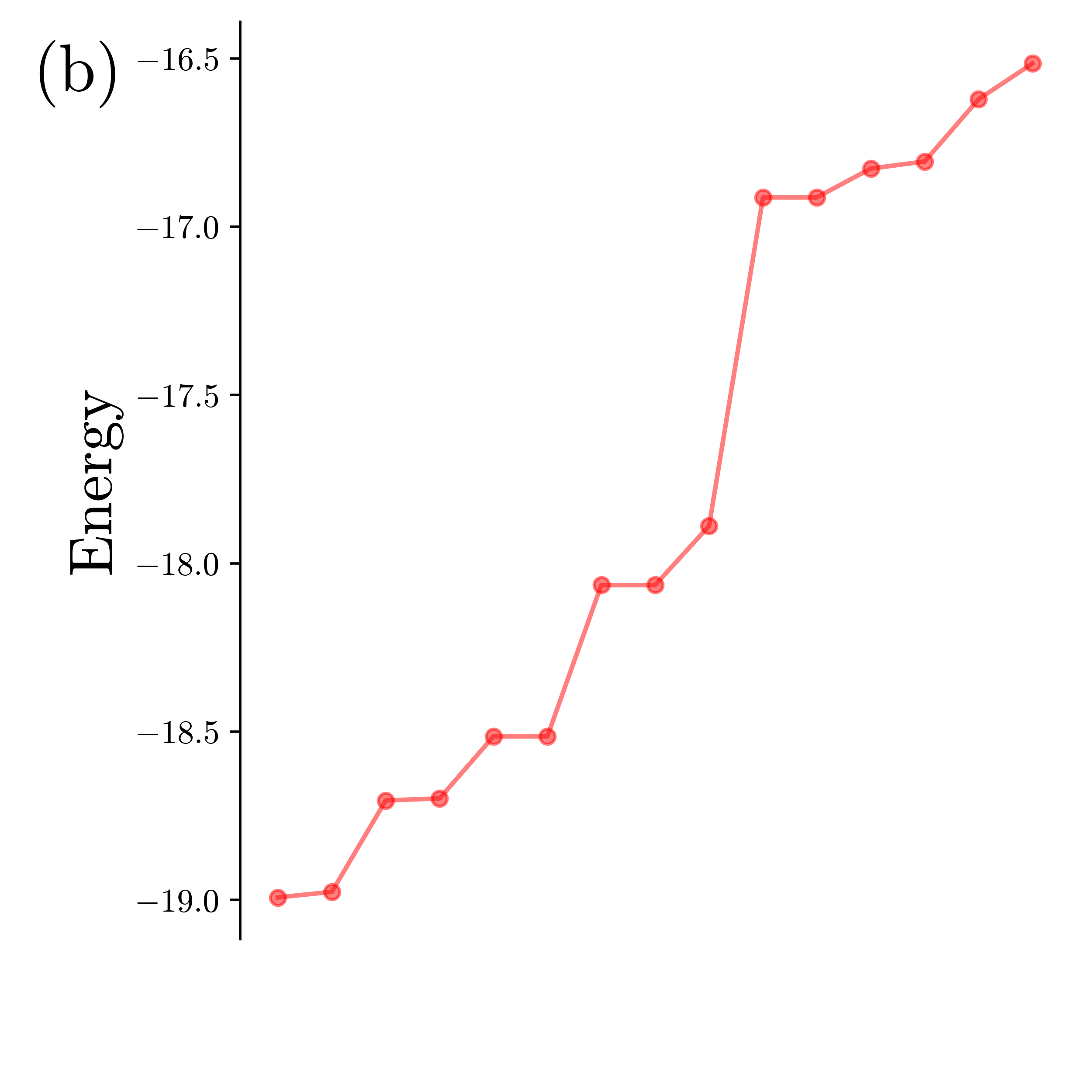}
\end{minipage} 
\begin{minipage}[ht]{0.38\linewidth}
\centering
\includegraphics[width=1\linewidth]{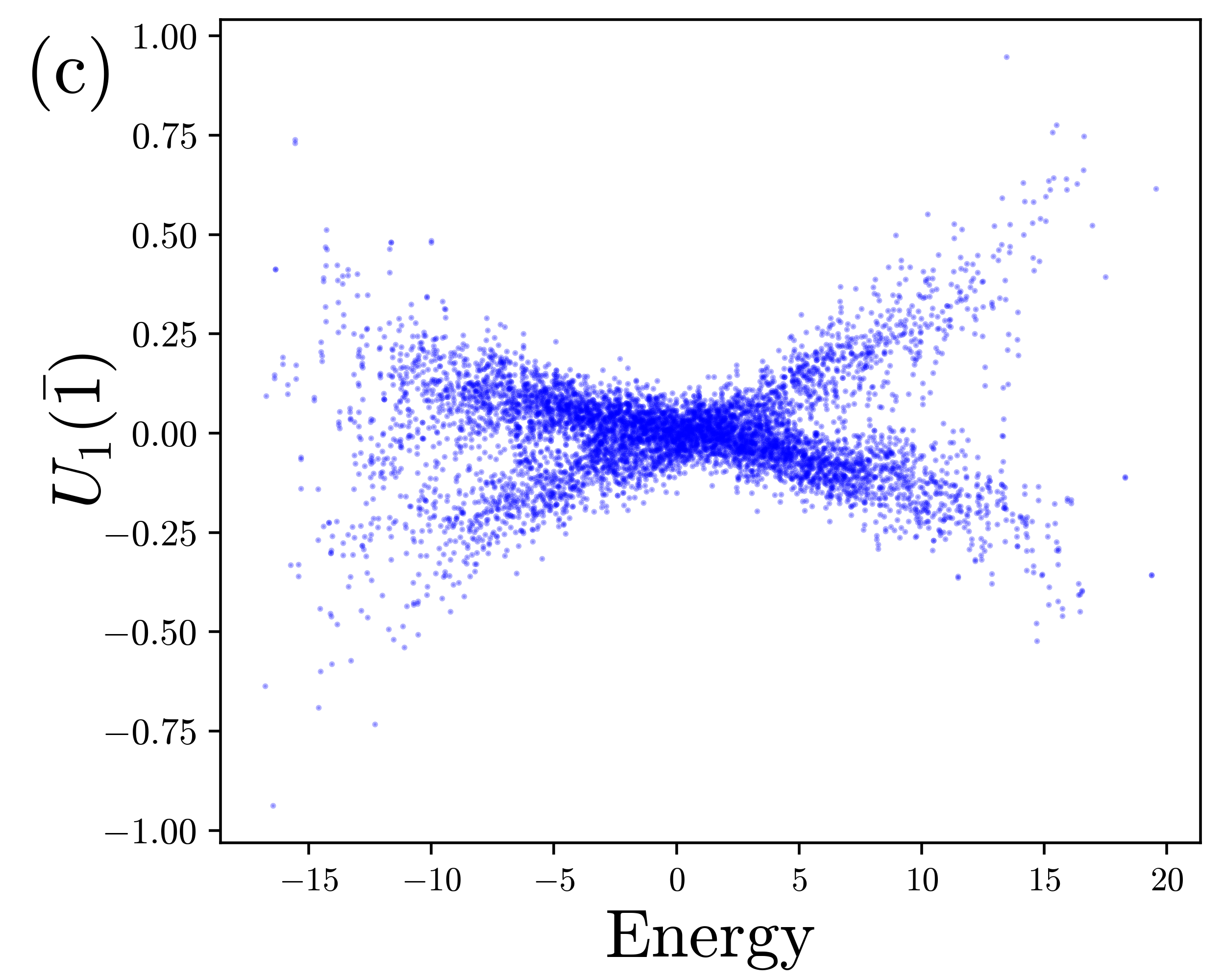}
\end{minipage} 
\caption{\footnotesize (a)(b) Part of energy spectra for $\mathbb{Z}_3\times \mathbb{Z}_3$-symmetry spin chain for $L=7$, $\lambda=0.1$. 
The coupling constants are uniformly distributed in $J_j^w\in[1.0+0.2i,1.1+0.2i]$, $J_j^x\in[0.9,1.0]$, $J_j^y\in[0.1,0.2]$, $J_j^z\in[0.2,0.3]$.
The degeneracy in the original Hamiltonian (\ref{3-Ham}) (a) is resolved by the perturbation (\ref{3-Ham-1}) (b).
(c) The expectation value of $U_1(\bar{1})$ for $L=8$, $\lambda=0.4$.
The expectations are separated into two sectors, and thus the ETH for $U_1(\bar{1})$ is not satisfied.
}\label{fig:spin_n3}
\end{figure}

Since the local operator $Z_j$ is not charged under $U_i$, a desired perturbation can be performed as
\begin{align}
    \tilde{H}_{N=3} := H_{N=3} + \lambda\sum_{j=1}^{L}Z_{j},
    \label{3-Ham-1}
\end{align}
and then the Hamiltonian is invariant only under the action of $U_i$.
As shown in Fig. \ref{fig:spin_n3}, the triple degeneracies of the energy spectrum are resolved by the perturbation, and the operator $U(\bar{1})=\prod_{j=2}^{L}Z_j$ does not satisfy the ETH.

\subsection{$(2+1)$-dimensional $\mathbb{Z}_2$ gauge theory}\label{sec:Z2 gauge}

Here we consider the (2+1)-dimensional $\mathbb{Z}_2$ lattice gauge theory defined on a $L_x\times L_y$ square lattice with the periodic boundary conditions.
In this model, the argument in Section \ref{sec:Projective} can be applied to the projective representation of $\mathbb{Z}_2$ electric one-form symmetry and ``time reversal'' symmetry.

\medskip

The Hamiltonian is given by \cite{Kogut:1974ag,Fradkin:1978th,Fradkin:1978dv}
\begin{align}
H_{\mathbb{Z}_2}=
- \sum_{\substack{\bm{r}}}\lambda_{\bm{r},xy} 
\sigma_{\bm{r},x}^3\sigma_{\bm{r}+\bm{e}_x,y}^3
\sigma_{\bm{r}+\bm{e}_y,x}^3\sigma_{\bm{r},y}^3
-
\lambda\sum_{\bm{r},j}\sigma_{\bm{r},j}^1,
\label{Z2-Hamiltonian}
\end{align}
where $\sigma_{\bm{r},j}^{1,2,3}$ denote the Pauli matrices acting on the link variable $(\bm{r},j)$, specified by the coordinate of vertices $\bm{r}$ and the direction $j=x,y$.
The coupling constants $\lambda_{\bm{r},xy}$ and $\lambda$ are real numbers.
Along the line of Section \ref{sec:Projective}, we can regard the term $\lambda\sum_{\bm{r},j}\sigma_{\bm{r},j}^1,$ as a perturbation term.
We can observe that for $\lambda=0$, the Hamiltonian $H_{\mathbb{Z}_2}$ is invariant under the ``time reversal'' symmetry\footnote{
This operation is not the time reversal symmetry in the usual sense, because it does not accompany the complex conjugation.
However, the complex conjugation does not affect the Hamiltonian $H_{\mathbb{Z}_2}$, and we just referred to this symmetry as "time reversal."
} represented by 
\begin{align}
\tilde{U}:= \prod_{\bm{r},j}\sigma_{\bm{r},j}^2.
\end{align}
This theory also enjoys the electric $\mathbb{Z}_2$ 1-form symmetry, and the spatial symmetry operators can be characterized by $H_1(T^2,\mathbb{Z}_2)=\mathbb{Z}_2\oplus\mathbb{Z}_2$~\cite{Roumpedakis:2022aik}.
The generators of $\mathbb{Z}_2$ correspond two independent symmetry operators corresponding to the $x$-cycle and $y$-cycle.

\medskip

Though the total Hilbert space of the system for the $L_x\times L_y$ lattice is $(2^{2L_xL_y})$-dimensional, we have to project it onto the physical Hilbert space. 
This is because there exist residual gauge redundancies, after the temporal gauge-fixing, which is analogous to the gauge $A_0(\bm{r})=0$ for the Maxwell theory.
Spatial gauge transformation is generated by the local operator
\begin{align}
    Q_{v}:=\prod_{\substack{b:\mbox{ \scriptsize spatial link}, b\ni v}}\sigma_{b}^1,\qquad
    v: \mbox{vertex},
\end{align}
which satisfies $Q_v^2=1$ and  $[H_{\mathbb{Z}_2},Q_v]=0$.
The physical Hilbert space is then obtained as
\begin{align}
\operatorname{span}\left\{ |\psi\rangle\;\big|\; Q_v|\psi\rangle = +|\psi\rangle ,\; {}^{\forall} v:\mbox{ vertices} \right\}.
\end{align}
This constraint can be regarded as the $\mathbb{Z}_2$ analog of the Gauss law $\nabla\cdot E|\psi\rangle_{\mathrm{phys}} = 0$ since we can write $Q_v=(\sigma_{\bm{r}-\bm{e}_x,x}^1)^{-1}\sigma_{\bm{r},x}^1(\sigma_{\bm{r}-\bm{e}_y,y}^1)^{-1}\sigma_{\bm{r},y}^1$.
After this projection, the expectation value of non-gauge invariant operators with respect to physical states $|\psi\rangle$ always vanishes.

\medskip

\begin{figure}[htbp]
\centering
\includegraphics[width=0.7\linewidth]{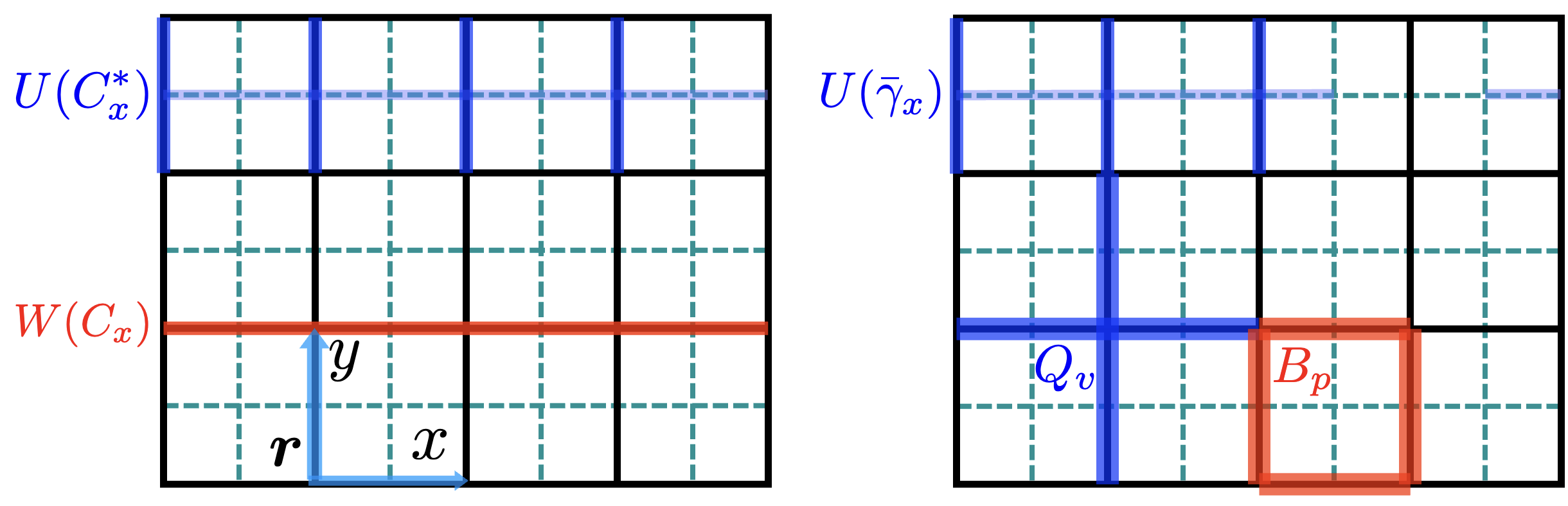}
\caption{\footnotesize Schematics for the action of the operators on the $4\times 3$ square lattice with the periodic boundary condition.
The lattice and the dual lattice are denoted by the solid and dashed lines, respectively.
The blue lines represent the action of $\sigma^1$, and the red lines indicates the action of $\sigma^3$.
}\label{fig:lattice}
\end{figure}

The Wilson and 't Hooft operators on the spatial directions are defined as~\cite{tHooft:1977nqb,Ukawa:1979yv,Shimazaki:2020byo}
\begin{align}
W(C)=\prod_{b\in C}\sigma_{b}^3, 
\qquad
U(C^*)=&\, \prod_{b^*\in C^*}\sigma_{b^{*}}^1, 
\end{align}
where, $C$ and $C^*$ are closed loops on the lattice and dual lattice, respectively (see Fig.~\ref{fig:lattice}).
Both $U(C^*)$ and $W(C)$ are gauge invariant operators since they commute with $Q_v$.

\medskip

The 't Hooft operator $U(C^*)$ satisfies
$
[H_{\mathbb{Z}_2}, U(C^*) ] =0,
$
and serves as the $\mathbb{Z}_2$ 1-form symmetry operator.
This operator is topological since 
continuous deformations of the path $C^*$ do not change the action of $U(C^*)$ on the physical states, i.e.,
$
U(C_1^*)|\psi\rangle = U(C_2^*)|\psi\rangle
\label{U-spatial}
$
if $C_1^*$ and $C_2^*$ are homotopically equivalent.
If a dual closed loop $C^*$ is topologically trivial, it follows that $U(C^*)|\psi\rangle=|\psi\rangle$.
The ``electric'' charge of the Wilson operator is measured by the 't Hooft operator $U(C^*)$.
Defining closed loops winding around the $x$-/$y$-cycle by $C_x$ and $C_y$ (and similarly the loops on the dual lattice by $C_x^*$ and $C_y^*$),
we see that the operators $W$ and $U$ satisfy 
\begin{align}\begin{split}
U(C_y^*) W(C_x) U^{-1}(C_y^*) = - W(C_x),\quad  U(C_x^*) W(C_y) U^{-1}(C_x^*) = - W(C_y),\\
U(C_x^*) W(C_x) U^{-1}(C_x^*) = + W(C_x),\quad  U(C_y^*) W(C_y) U^{-1}(C_y^*) = + W(C_y),
\end{split}\end{align}
which is indeed operator-realization of the electric $\mathbb{Z}_2$ 1-form symmetry~\cite{Pace:2023gye}.

\medskip

After these setups, we can explicitly observe that the symmetry operators satisfy $U(C_x^*)\tilde{U}=-\tilde{U}U(C_x^*)$ and $U(C_y^*)\tilde{U}=-\tilde{U}U(C_y^*)$.
This projective phase arises from the $\mathbb{Z}_2$ action by $\tilde{U}$ on the symmetry operator of the $\mathbb{Z}_2$ 1-form symmetry.
Introducing the one-dimensional operator 
\begin{align}
    \cZ(C^*):=\prod_{b^*\in C^*}(-1),
\end{align}
which generate $\mathbb{Z}_2$  ``1-form symmetry,'' we have
\begin{align}
    \tilde{U}U(C^*)\tilde{U}^{-1}=\cZ(C^*)U(C^*).
\end{align}
In fact, this relation is incorporated into a 2-group structure.

\medskip

\begin{figure}[htbp]
\centering
\begin{minipage}[ht]{0.4\linewidth}
\centering
\includegraphics[width=1.\linewidth]{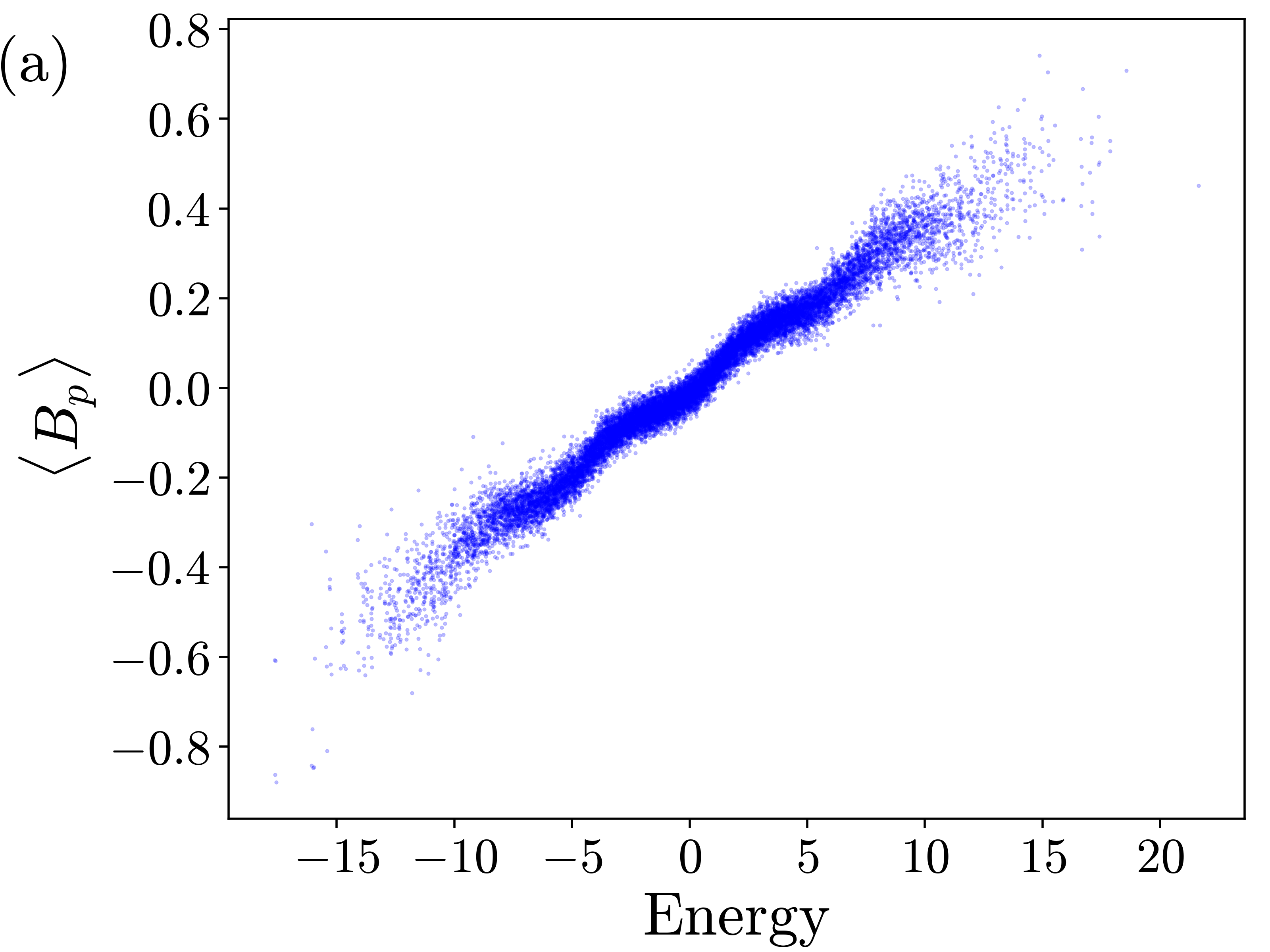}
\end{minipage} 
\hspace{0pt}
\begin{minipage}[ht]{0.4\linewidth}
\centering
\includegraphics[width=1.\linewidth]{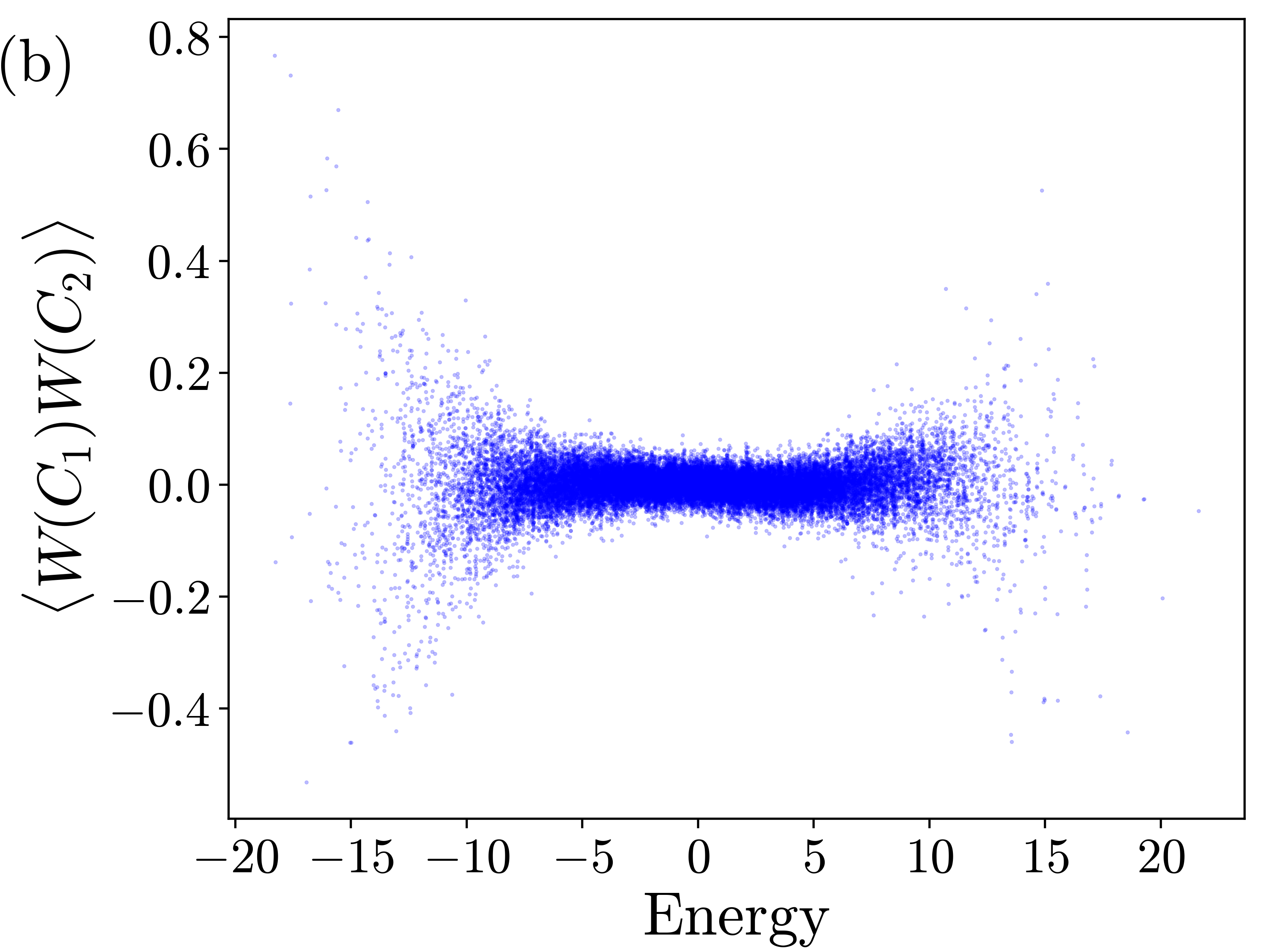}
\end{minipage} 
\caption{\footnotesize 
(a)(b) The expectation values of the plaquette operator $B_p:=\prod_{i\in p:\mathrm{plaquette}}\sigma^3_i$ and a double insertion of the Wilson operators $W(C_1)W(C_2)$ with respect to the energy eigenstates for the $\mathbb{Z}_2$ gauge theory.
The ETH for both of the operators are satisfied while the operator $W(C_1)W(C_2)$ is a 1-dimensional non-local operator.
}\label{fig:Z2-gauge}
\end{figure}

The perturbation $H_1=\sum_{\bm{r},j}\sigma_{\bm{r},j}^1$ then lift the degeneracy, and 
this suffices to show the breakdown of the ETH as shown in Fig.~\ref{fig:demo}, where the numerical calculation are performed for the $5\times 3$ lattice, and the coupling constants $\lambda_{\bm{r},xy}$ are uniformly chosen from $[1.0,1.1]$ and the parameter is set to $\lambda =0.6$.
On the other hand, the ETH holds for other operators such as the plaquette operator $B_p:=\prod_{i\in p:\mathrm{plaquette}}\sigma^3_i$ and a double insertion of the Wilson operators $W(C_1)W(C_2)$ (Fig.~\ref{fig:Z2-gauge}).
We stress that the nonlocality of the operator does not immediately lead to the breakdown of the ETH since the non-local operator $W(C_1)W(C_2)$ satisfy the ETH.

\section{Conclusion and discussion}\label{sec:conclusion}

In this paper, we have shown that the the one of the sufficient conditions for the ETH-violation is satisfied if we consider perturbations that break the symmetry with nontrivial projective phases.
Following this treatment, we just have to suppose the following:
1) the unperturbed Hamiltonian exhibits a $\mathbb{Z}_N\times\mathbb{Z}_N$ ($p$-form, in general) symmetry with a nontrivial projective representation; 2) the symmetry operator corresponding to one of the $\mathbb{Z}_N$ symmetry can be divided as $U_m(\tilde{C})=U_m(\gamma)U_{m}(\bar{\gamma})$ with open manifolds $\gamma$ and $\bar{\gamma}$; 3) $\langle U_m(\gamma)\rangle_{\mathrm{mc}}^{\delta E}\neq 0$.
Under these assumptions, the ETH for the $(d-p)$-dimensional operator $U_m(\gamma)$ or $U_m(\bar{\gamma})$ is always violated after perturbing the Hamiltonian by $\lambda\sum_{j}W_q(j)$ with the scaling (\ref{scale-supposition}).
Although conditions ii) in Section \ref{sec:p-ETH} (and in \cite{Fukushima:2023swb}) require information about each eigenstate in the middle of the spectrum a priori,  tractable conditions 1), 2), and 3) to such eigenstates lead to the same conclusion.

\medskip

We also performed numerical calculations for $(1+1)$-dimensional $\mathbb{Z}_2\times\mathbb{Z}_2$-symmetric/$\mathbb{Z}_3\times\mathbb{Z}_3$-symmetric spin chains, and $(2+1)$-dimensional $\mathbb{Z}_2$ gauge theory.
All of the models have indeed the 1-dimensional ETH-violating operators, and their mechanisms are boiled down to the general discussion above.
We can thus conclude that our treatment indeed results in the ETH-violation for these concrete examples.

\medskip

As an outlook, the application of our formulation to other groups would be possible since essential mixture of the symmetry sectors are common to the $\mathbb{Z}_N\times\mathbb{Z}_N$ case.
It allows us to find the breakdown of the ETH for broader class of systems with higher-form symmetry including lattice gauge theories.
 In \cite{Lashkari:2016vgj}, the local ETH leads to the subsystem ETH for the 0-dimensional subsystems. On the other hand, Many 0-form symmetries nest in general 2-dimensional CFTs, and thus the result in this paper means the violation of the ETH for the 1-dimensional operators. This observation motivate us to consider consequences for the subsystem ETH for 1-dimensional subsystems in the context of CFTs.
It would also be interesting to explore consequences of the mixed 't Hooft anomaly on thermalization processes in our various gauge theories such as Yang-Mills theories \cite{Hayata:2023puo,Biswas:2022env}.
We hope our results shed light on non-equilibrium dynamics for various quantum field theories and quantum many-body systems.


\subsection*{Acknowledgments}

We are very grateful to R. Hamazaki for helpful comments on the manuscript.
The work of O.\,F.\ was supported by Grant-in-Aid for JSPS Fellows No.~21J22806.

%


\bibliographystyle{utphys}
\bibliography{proj_ETH}

\end{document}